\documentclass[]{raa}            

\usepackage{graphicx,times}             
\usepackage{natbib}
\usepackage{amssymb,amsmath}
\bibpunct{(}{)}{;}{a}{}{,}

\usepackage{tablefootnote}

\usepackage[a4paper=true,pagebackref=true]{hyperref}
\hypersetup{colorlinks = true, linkcolor = green, anchorcolor = red, citecolor = blue, filecolor = red, pagecolor = red, urlcolor = red}

\begin{document}

   \title{On the correctness of orbital solutions obtained from a small set of points. Orbit of HIP~53731.
}

   \volnopage{Vol.0 (20xx) No.0, 000--000}      
   \setcounter{page}{1}          

   \author{A. Mitrofanova
      \inst{}
   \and V. Dyachenko
      \inst{}
   \and A. Beskakotov
      \inst{}
   \and Yu. Balega
      \inst{}
   \and A. Maksimov
   	  \inst{}
   \and D. Rastegaev
   	  \inst{}
   \and S. Komarinsky
   	  \inst{}
   }

   \institute{Special Astrophysical Observatory of the Russian Academy of Sciences, Nizhnij Arkhyz, Russia 369167; {\it arishka670a@mail.ru}\\
\vs\no
   {\small Received~~20xx month day; accepted~~2020~~May 31}}

\abstract{HIP~53731 is a binary consisting of stars of the spectral types K0 and K9. Orbit of this object was constructed previously by \citet{cve16} and improved by \citet{tok19orb}. It should be noted that there is an 180$^\circ$ ambiguity in the position angles of some published measurements. Speckle interferometric observations were obtained in 2007-2020 (21 measurements) at the 6-m telescope of the SAO RAS (BTA) by the authors of this article. The analysis of new data together with previously published ones made it possible to construct the accurate orbit of HIP~53731 and to halve the already known values of the orbital period of the system. As a result of the study, the mass sum, the masses of each component and their spectral types were determined by two independent methods. According to the qualitative classification of orbits, the orbital solution has grade 2 - “good” (observations cover more than half of the orbital period and correspond to different phases). 
}

   \authorrunning{A. Mitrofanova, V. Dyachenko, A. Beskakotov et al.}            
   \titlerunning{Orbit of HIP~53731.}  

   \maketitle

%
%
\twocolumn

\section{Introduction}           
\label{sect:intro}

Currently, a significant part of the orbital solutions found from speckle interferometric data for binaries is obtained using a small number of measurements. This may be due to the lack of observations for a particular system, as well as to large orbital periods of such objects. An additional factor is the absence or inaccurate determination of the positions of the secondaries, which is difficult to identify due to a lack of observational data. Therefore orbital parameters of objects and mass sums could be determined incorrectly. One of the solutions to this problem is the long-term monitoring of binaries, which allows for covering big part of the orbit with measurements (it depends on the orbital period). In this paper, we studied the binary HIP~53731 (HD~95175, $V_{mag}=8.85^{m}$ \citep{bid85}). The system under study, according to \citet{cve16}, consists of two main sequence stars of spectral types K0 and K9 with masses $\mathfrak{M}_{A} = 0.9~ \mathfrak{M}_{\odot}$ and $\mathfrak{M}_{B} = 0.48~ \mathfrak{M}_{\odot}$, and the total mass of the system and dynamical parallax are $\mathfrak{M}_{tot} = 1.72~ \pm 0.4 \mathfrak{M}_{\odot}$ and $\pi_{dyn} = 29.33 \pm 3.03$ mas. Positional parameters published earlier and obtained in this study are presented in Table \ref{tab1}. Speckle interferometric observations and the image reduction procedure are discussed in Section \ref{sect:Obs}, Section \ref{sect:orbit} is dedicated to the process of orbit construction and determination of the fundamental parameters of HIP~53731, results are discussed in Section \ref{sect:Dis}.

\section{Observations and data reduction}
\label{sect:Obs}

Speckle interferometric observations of HIP~53731 were carried out at the Big Telescope Alt-azimuth (BTA) of the Special Astrophysical Observatory of the Russian Academy of Sciences (SAO RAS) from 2007 to 2020 using a speckle interferometer \citep{maks09} based on EMCCD detectors PhotonMAX-512B (until 2010), Andor iXon+ X-3974 (2010-2014) and Andor iXon Ultra 897 (since 2015). Speckle images were obtained under good weather conditions with seeing about 1\arcsec-2\arcsec. Speckle interferograms were recorded with exposure time of 20 milliseconds, the standard series consisted of 1940 (until 2010) and 2000 images. Following interference filters were used (central wavelength $\lambda$ / bandpass $\Delta\lambda$): 550/20, 600/40 and 800/100 nm. 

Positional parameters and magnitude differences were determined on the basis of the analysis of the power spectrum and the autocorrelation function of the speckle interferometric series described in \citet{bali02} and \citet{pluz05}. The reconstruction of the position of the secondary was carried out by the bispectrum method \citep{lohm83}. The log of observations, positional parameters and the magnitude differences are presented in Table \ref{tab1}: epoch of observations in fractions of the Besselian year; telescope; $\lambda$/$\Delta\lambda$; $\theta$ is the position angle; $\rho$ is the separation between the two stars; $\Delta m$ is the magnitude difference and references. The formal errors of $\Delta m$ corresponding to the method of model selection are presented in Table \ref{tab1}. Wherein, an analysis of the magnitude differences from the data obtained in different epochs shows that the actual measurement accuracy is about 0.1 mag.

\begin{table*}
	\begin{center}
	\caption[]{Positional Parameters and Magnitude Differences.}\label{tab1}
	\begin{tabular}{|c|c|c|c|c|c|c|}
	\hline\noalign{\smallskip}
Epoch & Telescope & $\lambda$/$\Delta\lambda$, nm & $\theta^{\circ}$ & $\rho$, mas & $\Delta m$, mag & Reference \\
	\hline\noalign{\smallskip}
1991.25 & \textit{Hipparcos} &  & 291.0 & 287.0 &  & \citet{hip} \\
2000.1460 & 3.5-m WIYN & 648/41 & $275.8 \pm 1$ & $184 \pm 3$ &  & \citet{hor02} \\
2001.2733 & BTA & 600/30 & $268.9 \pm 0.6$ & $177 \pm 4$ &  & \citet{bali06} \\
2001.2733 & BTA & 750/35 & $268.9 \pm 0.7$ & $178 \pm 4$ &  & \citet{bali06} \\
2002.2542 & BTA & 750/35 & $260.0 \pm 0.6$ & $144 \pm 2$ &  & \citet{bali13} \\
2005.2323 & BTA & 800/110 & $118.8 \pm 1.1$ & $139 \pm 3$ &  & \citet{bali13} \\
2006.3745 & BTA & 545/30 & $107.8 \pm 1.2$ & $180 \pm 4$ &  & \citet{bali13} \\
2007.9019 & BTA & 600/30 & $277.8 \pm 0.1$ & $191 \pm 1$ & $2.05 \pm 0.01$ & this work \\
2008.9559 & BTA & 550/20 & $270.4 \pm 1$ & $175 \pm 1$ & $2.55 \pm 0.02$ & this work \\
2009.0954 & BTA & 600/40 & $268.7 \pm 0.1$ & $175 \pm 1$ & $2.29 \pm 0.01$ & this work \\
2009.2645 & BTA & 600/40 & $267.7 \pm 0.1$ & $169 \pm 1$ & $1.64 \pm 0.01$ & this work \\
2010.1601 & BTA & 800/100 & $259.3 \pm 0.1$ & $141 \pm 1$ & $1.45 \pm 0.01$ & this work \\
2010.3416 & 2.1-m OAN & 630/120 & $55.5 \pm 12.9$ & $160 \pm 30$ &  & \citet{orl15} \\
2011.1351 & BTA & 800/100 & $240.1 \pm 0.3$ & $83 \pm 1$ & $1.53 \pm 0.02$ & this work \\
2011.9486 & BTA & 800/100 & $343.9 \pm 0.2$ & $34 \pm 1$ & $1.51 \pm 0.01$ & this work \\
2013.3221 & BTA & 550/20 & $296.4 \pm 0.1$ & $150 \pm 1$ & $2.24 \pm 0.01$ & this work \\
2014.1193 & BTA & 800/100 & $289.5 \pm 0.1$ & $179 \pm 1$ & $1.46 \pm 0.01$ & this work \\
2014.9301 & BTA & 550/20 & $283.4 \pm 0.2$ & $188 \pm 1$ & $2.11 \pm 0.01$ & this work \\
2015.9703 & BTA & 800/100 & $274.8 \pm 0.1$ & $189 \pm 1$ & $1.55 \pm 0.01$ & this work \\
2016.1331 & 4.1-m SOAR & 788/132 & $274.0 \pm 0.3$ & $193.5 \pm 0.8$ & 1.5 & \citet{tok18} \\ 
2016.8847 & BTA & 800/100 & $268.8 \pm 0.1$ & $174 \pm 1$ & $1.56 \pm 0.01$ & this work \\
2017.9227 & BTA & 800/100 & $258.9 \pm 0.1$ & $141 \pm 1$ & $1.46 \pm 0.01$ & this work \\
2017.9227 & BTA & 800/100 & $259.0 \pm 0.1$ & $141 \pm 1$ & $1.46 \pm 0.01$ & this work \\
2017.9227 & BTA & 800/100 & $258.9 \pm 0.1$ & $141 \pm 1$ & $1.47 \pm 0.01$ & this work \\
2017.9227 & BTA & 800/100 & $258.9 \pm 0.1$ & $142 \pm 1$ & $1.5 \pm 0.01$ & this work \\
2018.1811 & 4.1-m SOAR & 824/170 & $255.2 \pm 0.8$ & $134.5 \pm 0.8$ & 1.5 & \citet{tok19} \\
2018.3214 & BTA & 800/100 & $253.7 \pm 0.1$ & $124 \pm 1$ & $1.49 \pm 0.01$ & this work \\
2019.0476 & BTA & 800/100 & $235.4 \pm 0.1$ & $76 \pm 1$ & $1.47 \pm 0.02$ & this work \\
2019.2744 & BTA & 550/20 & $222.4 \pm 0.1$ & $61 \pm 1$ & $1.9 \pm 0.04$ & this work \\
2019.2744 & BTA & 800/100 & $221 \pm 0.2$ & $64 \pm 1$ & $1.27 \pm 0.04$ & this work \\
2020.3611 & BTA & 550/20 & $306 \pm 0.1$ & $103 \pm 1$ & $2.13 \pm 0.01$ & this work \\
	\noalign{\smallskip}\hline
\end{tabular}
\end{center}
\end{table*}

\section{Orbit Construction}
\label{sect:orbit}

\begin{figure}
	\centering
	\includegraphics[width=7.5cm, angle=0]{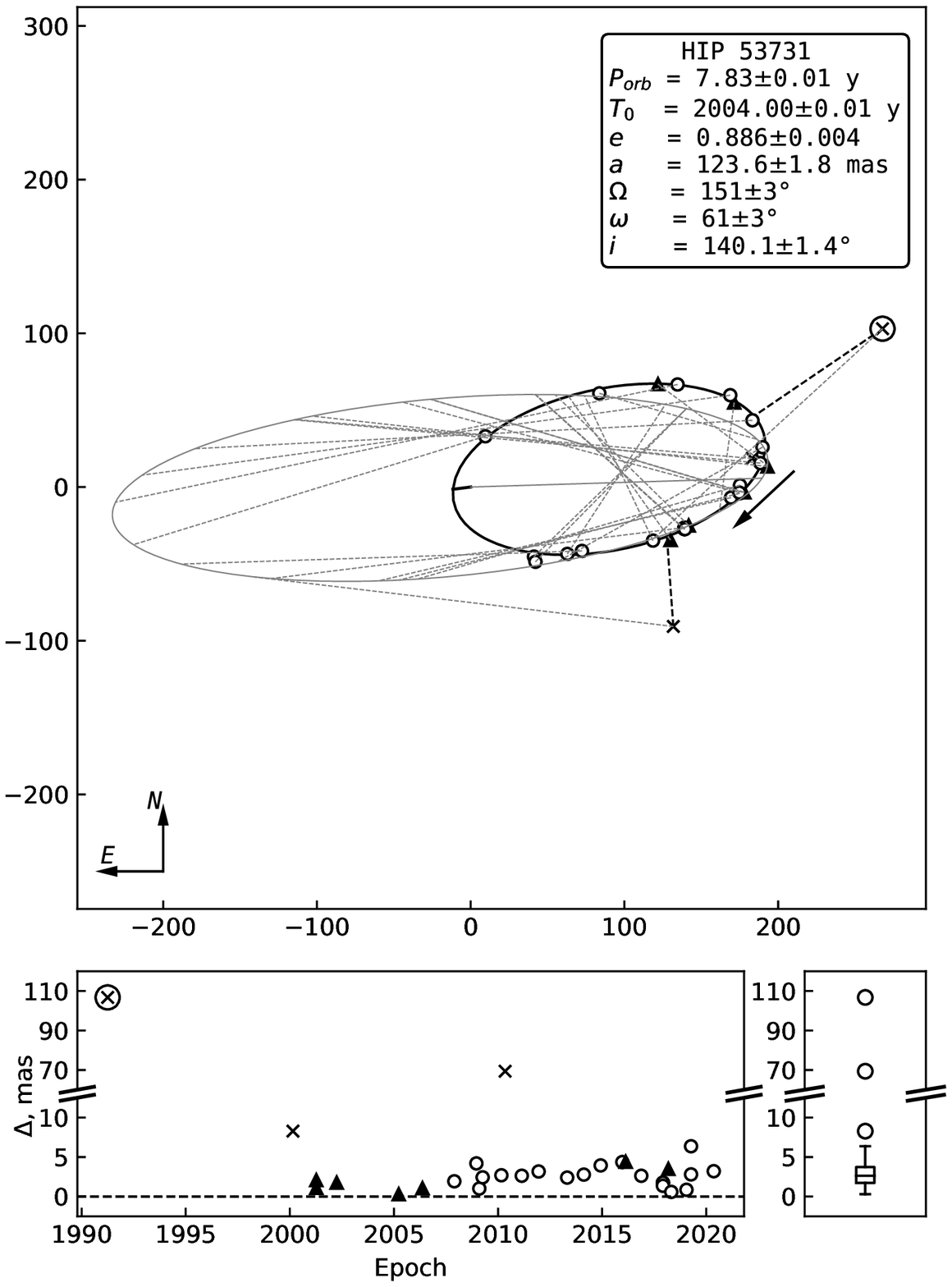}
	\caption{Orbital solutions for HIP~53731. The orbit by \citet{tok19orb} is marked with gray and the orbit constructed in this work is black. Triangles correspond to the published data; open circles - data obtained in this study; crosses - data with large residuals; the cross placed in a large circle is the first measurement for system. The arrow shows the direction of motion of the secondary. $\Delta$ are residuals showing the angular distance between the observed and modelled value. The dashed line on the residuals plot indicates the orbital solution.}
	\label{fig1}
	\end{figure}

Preliminary estimates of the orbital parameters were calculated using the Monet method \citep{mon77}. The final orbit was constructed using the ORBIT software package \citep{tok92}. Depending on the values of residuals and deviations from the orbital solution, the corresponding weights were selected for each measurement. Measurements by \citet{hip}, \citet{hor02} and \citet{orl15} have the largest residuals (as it is shown in Figure \ref{fig1}), consequently less weight was set to them.

When constructing the orbit of HIP~53731, ambiguities in the positions of published measurements were found, which is probably due to the incorrect reconstruction of the position of the secondary or its absence. As a result, the position angles of the following measurements were changed by $\pm 180^\circ$: 2005.2323 and 2006.3745 \citep{bali13} and 2010.3416 \citep{orl15}. Two orbits of HIP~53731 constructed by \citet{tok19orb} and in this study are presented in Figure \ref{fig1}. The measurement residuals with respect to the new orbital solution are $4^\circ$ by $\theta$ and 18 mas by $\rho$. However, they were overestimated due to the significant contribution of points that obviously don't match to the model solution (in Figure \ref{fig1}, these are marked with crosses). Real estimates of residuals for $\rho$ and $\theta$ are 2 mas and $0.8^\circ$, respectively.

Table \ref{tab2} presents our measurements of the orbital parameters of the system and previously published ones. The columns are: the orbital period, the epoch of passing the periastron, the eccentricity of the orbit, the semimajor axis, the longitude of the ascending node, the argument of the periastron, the inclination of the orbit and the references to the publications.

\begin{table*}
\begin{center}
\caption[]{Orbital Parameters of HIP~53731.}\label{tab2}
\begin{tabular}{|c|c|c|c|c|c|c|c|}
  \hline\noalign{\smallskip}
$P_{orb}$, year & $T_{0}$, year & $e$ & $a$, mas & $\Omega$, $^\circ$ &  $\omega$, $^\circ$ & $i$, $^\circ$ & Reference \\
\hline\noalign{\smallskip}
16.244 & 2003.412 & 0.150 & 202.6 & 95.3 & 256.6 & 115.8 & \citet{cve16} \\ 
$\pm 0.682$ & $\pm 0.682$ & $\pm 0.041$ & $\pm 10.9$ & $\pm 2.8$ & $\pm 15.8$ & $\pm 1.6$ & \\ 
\hline\noalign{\smallskip}
15.83 & 2000.92 & 0.095 & 214 & 95.1 & 192.2 & 105.8 & \citet{tok19orb} \\
\hline\noalign{\smallskip}
7.83 & 2004.00 & 0.886 & 123.6 & 151 & 61 & 140.1 & this \\
$\pm 0.01$& $\pm 0.01$ & $\pm 0.004$ & $\pm 1.8$ & $\pm 3$ & $\pm 3$ & $\pm 1.4$ & work \\
  \noalign{\smallskip}\hline
\end{tabular}
\end{center}
\end{table*} 

A comparison of orbital solutions shows, that the orbit of HIP~53731 constructed in this work fits the observational data much better, than the orbits by \citet{cve16} and by \citet{tok19orb}, obtained from a small amount of observational data. Also, mass sum, absolute magnitudes of components, their spectral types and masses were determined using two independent methods and are presented in Table \ref{tab3}. Both \textit{Hipparcos} \citep{hippi} and \textit{Gaia} \citep{gapi} parallaxes and orbital parameters were used in the first method. This method allows for calculation of mass sum via the \textit{Kepler's} law:

\begin{equation}
\sum \mathfrak{M}=\frac{(a/\pi)^{3}}{P_{orb}^{2}},
\label{eq1}
\end{equation}

and the uncertainty is calculated using

\begin{equation}
\begin{gathered}
\sigma(\mathfrak{M})=\\
\sqrt{\frac{9(\sigma_{\pi})^2}{\pi^2}+\frac{9(\sigma_{a})^2}{a^2}+\frac{4(\sigma_{P_{orb}})^2}{P_{orb}^2}}*\mathfrak{M}.
\end{gathered}
\label{eq2}
\end{equation}

The second method allows for obtaining of the masses of components via the Pogson's relation. The magnitude of the object in the V band \citep{bid85} and the average magnitude difference ($\Delta m_{550} = 2.19^{m} \pm 0.10^{m}$ from Table \ref{tab1}) together with parallaxes ($\pi_{\textit{Hip}} = 26.35 \pm 1.29$ mas and $\pi_{\textit{Gaia}} = 31.0803 \pm 0.6137$ mas) were used. The work by \citet{pec13} was applied to match the calculated absolute magnitudes of the components with spectral types and masses.

\begin{table*}
	\begin{center}
		\caption[]{Comparison of Fundamental Parameters.}\label{tab3}
		\begin{tabular}{|c|c|c|c|c|c|c|c|c|}
			\hline\noalign{\smallskip}
	& Parallax & $M_{V,A}$, m & $Sp_{A}$ & $\mathfrak{M}_{A}$, $\mathfrak{M}_{\odot}$ & $M_{V,B}$, m &  $Sp_{B}$ & $\mathfrak{M}_{A}$, $\mathfrak{M}_{\odot}$ & $\sum \mathfrak{M}$, $\mathfrak{M}_{\odot}$ \\
	\hline\noalign{\smallskip}
\citet{cve16} & \textit{Hipparcos} & $5.99 \pm 0.12$ & K0 & 0.90 & $8.57 \pm 0.77$ & K9 & 0.48 & $1.72 \pm 0.40$ \\ 
	\hline\noalign{\smallskip}
This & \textit{Hipparcos} & $6.11 \pm 0.10$ & K2 & 0.78 & $8.30 \pm 0.14$ & K7 & 0.63 & $1.68 \pm 0.26$ \\ \cline{2-9}
work & \textit{Gaia} & $6.47 \pm 0.10$ & K3 & 0.75 & $8.66 \pm 0.14$ & K9 & 0.56 & $1.03 \pm 0.07$ \\
  \noalign{\smallskip}\hline
\end{tabular}
\end{center}
\end{table*}

\section{Discussion}
\label{sect:Dis}
An analysis of speckle interferometric data obtained at the 6-m telescope of the SAO RAS from 2007 to 2020 made it possible to halve the previously known value of the orbital period of HIP~53731. It should be noted that the residuals of positional parameters are small, which indicates the high-precision of the orbit. This fact indicates a high accuracy of new orbital parameters and the justification for long-term monitoring of such objects carried out in the group of high-resolution methods in astronomy of the SAO RAS. As a result, the orbital solutions by \citet{cve16} and by \citet{tok19orb} are very different from one, presented in this work, because they were obtained using small number of measurements, some of which have $\pm 180^\circ$ ambiguities. 

The mass sum of the HIP~53731 components was determined with an accuracy of 15\% (using the \textit{Hipparcos} parallax) and 8\% (using the \textit{Gaia} parallax). The masses of the components obtained by the second method in this study are consistent with the mass sums calculated by the first method. The values obtained using \textit{Hipparcos} parallax agree better with each other. The masses obtained using \textit{Gaia} parallax in this work are less consistent with each other. The reason is probably \textit{Gaia} parallax, so we are looking forward to the new data release of this mission. The proximity of the new parameters to the previous values is explained by the fact that \citet{cve16} used the magnitude difference from the \textit{Hipparcos} catalog together with the table of star parameters from the book by \citet{gra05}, as well as erroneous values of both the orbital period and the semimajor axis. 

The classification of the obtained orbital solutions was carried out using the qualitative grade of \citet{wor83}. The orbit of HIP~53731 is ''good'' (Grade 2) - the observations correspond to different phases and cover more than half of the orbital period, which allows for fitting of orbit accurately enough.

\begin{acknowledgements}
The reported study was funded by RFBR, project number 20-32-70120. The work was performed as part of the government contract of the SAO RAS approved by the Ministry of Science and Higher Education of the Russian Federation. This work has made use of data from the European Space Agency (ESA) mission \textit{Gaia} (\url{https://www.cosmos.esa.int/gaia}), processed by the \textit{Gaia} Data Processing and Analysis Consortium (DPAC, \url{https://www.cosmos.esa.int/web/gaia/dpac/consortium}). Funding for the DPAC has been provided by national institutions, in particular the institutions participating in the \textit{Gaia} Multilateral Agreement. This research has made use of the SIMBAD database, operated at CDS, Strasbourg, France.  
\end{acknowledgements}

\bibliographystyle{raa}
\bibliography{53731}

\begin{thebibliography}{21}
\providecommand\natexlab[1]{#1}
\providecommand\JournalTitle[1]{#1}

\bibitem[{Balega} {et~al.}(2006)]{bali06}
{Balega}, I.~I., {Balega}, A.~F., {Maksimov}, E.~V., {et~al.} 2006, Bulletin of
  the Special Astrophysics Observatory, 59, 20

\bibitem[{Balega} {et~al.}(2013)]{bali13}
{Balega}, I.~I., {Balega}, Y.~Y., {Gasanova}, L.~T., {et~al.} 2013,
  Astrophysical Bulletin, 68, 53

\bibitem[{Balega} {et~al.}(2002)]{bali02}
{Balega}, I.~I., {Balega}, Y.~Y., {Hofmann}, K.-H., {et~al.} 2002, \aap, 385,
  87

\bibitem[{Bidelman}(1985)]{bid85}
{Bidelman}, W.~P. 1985, \apjs, 59, 197

\bibitem[{Cvetkovi{\'c}} {et~al.}(2016)]{cve16}
{Cvetkovi{\'c}}, Z., {Pavlovi{\'c}}, R., \& {Ninkovi{\'c}}, S. 2016, \aj, 151,
  83

\bibitem[{ESA}(1997)]{hip}
{ESA}, ed. 1997, ESA Special Publication, Vol. 1200, {The HIPPARCOS and TYCHO
  catalogues. Astrometric and photometric star catalogues derived from the ESA
  HIPPARCOS Space Astrometry Mission}

\bibitem[{Gaia Collaboration}(2018)]{gapi}
{Gaia Collaboration}. 2018, VizieR Online Data Catalog, I/345

\bibitem[{Gray}(2005)]{gra05}
{Gray}, D.~F. 2005, {The Observation and Analysis of Stellar Photospheres}

\bibitem[{Horch} {et~al.}(2002)]{hor02}
{Horch}, E.~P., {Robinson}, S.~E., {Meyer}, R.~D., {et~al.} 2002, \aj, 123,
  3442

\bibitem[{Lohmann} {et~al.}(1983)]{lohm83}
{Lohmann}, A.~W., {Weigelt}, G., \& {Wirnitzer}, B. 1983, \ao, 22, 4028

\bibitem[{Maksimov} {et~al.}(2009)]{maks09}
{Maksimov}, A.~F., {Balega}, Y.~Y., {Dyachenko}, V.~V., {et~al.} 2009,
  Astrophysical Bulletin, 64, 296

\bibitem[{Monet}(1977)]{mon77}
{Monet}, D.~G. 1977, \apj, 214, L133

\bibitem[Orlov \& Voitsekhovich(2015)]{orl15}
Orlov, V., \& Voitsekhovich, V. 2015, Revista mexicana de astronom{\'\i}a y
  astrof{\'\i}sica, 51, 65

\bibitem[{Pecaut} \& {Mamajek}(2013)]{pec13}
{Pecaut}, M.~J., \& {Mamajek}, E.~E. 2013, \apjs, 208, 9

\bibitem[{Pluzhnik}(2005)]{pluz05}
{Pluzhnik}, E.~A. 2005, \aap, 431, 587

\bibitem[{Tokovinin}(1992)]{tok92}
{Tokovinin}, A. 1992, in Astronomical Society of the Pacific Conference Series,
  Vol.~32, IAU Colloq. 135: Complementary Approaches to Double and Multiple
  Star Research, ed. H.~A. {McAlister} \& W.~I. {Hartkopf}, 573

\bibitem[{Tokovinin}(2019)]{tok19orb}
{Tokovinin}, A. 2019, INTERNATIONAL ASTRONOMICAL UNION COMMISSION G1 (BINARY
  AND MULTIPLE STAR SYSTEMS) DOUBLE STARS INFORMATION CIRCULAR, ed. J.A. Docobo
  and J.F. Ling, 198

\bibitem[{Tokovinin} {et~al.}(2018)]{tok18}
{Tokovinin}, A., {Mason}, B.~D., {Hartkopf}, W.~I., {Mendez}, R.~A., \&
  {Horch}, E.~P. 2018, \aj, 155, 235

\bibitem[{Tokovinin} {et~al.}(2019)]{tok19}
{Tokovinin}, A., {Mason}, B.~D., {Mendez}, R.~A., {Horch}, E.~P., \&
  {Brice{\~n}o}, C. 2019, \aj, 158, 48

\bibitem[{van Leeuwen}(2007)]{hippi}
{van Leeuwen}, F. 2007, \aap, 474, 653

\bibitem[{Worley} \& {Heintz}(1983)]{wor83}
{Worley}, C.~E., \& {Heintz}, W.~D. 1983, Publications of the U.S.~Naval
  Observatory Second Series, 24

\end{thebibliography}

\label{lastpage}

\end{document}